\documentclass[aps,pra,twocolumn,superscriptaddress,nofootinbib]{revtex4-2}

\usepackage{amsmath}
\usepackage{mathtools}
\usepackage{amssymb}
\usepackage{physics}
\usepackage{color}
\usepackage{graphicx}
\usepackage{import}
\usepackage{tabularx}
\usepackage{dcolumn} 
\usepackage{bm}      
\usepackage{array}   
\usepackage{amsfonts}
\usepackage[T1]{fontenc}

\usepackage[english]{babel}
\usepackage[bookmarks=false,colorlinks,citecolor=red]{hyperref}

\usepackage[utf8]{inputenc}
\usepackage[T1]{fontenc}
\usepackage{mathptmx}

\include{MyCommand}

\DeclareUnicodeCharacter{2212}{-}

\makeatletter
\newcommand*{\addFileDependency}[1]{
  \typeout{(#1)}
  \@addtofilelist{#1}
  \IfFileExists{#1}{}{\typeout{No file #1.}}
}
\makeatother

\include{MyCommand}

\begin{document} 


\title{ Quantum signatures and semiclassical limitations in the transmission of Fock states
} 




\author{D~.~J.~ Nader}
 \affiliation{Department of Optics, Faculty of Science, Palack\'y University, Olomouc 77146, Czech Republic}

 

\begin{abstract}
Transmission through potential barriers is a fundamental problem in quantum mechanics.
While semiclassical methods can approximate certain aspects of transmission, they fail to capture the intrinsically quantum interference associated with Wigner-function negativity.
We numerically study the transmission of displaced Fock states through an inverted-oscillator barrier, with and without a Kerr nonlinearity that offers a potential route to experimental realization.
These states allow only an approximate classical description, since their characteristic Wigner-function negativity is absent in phase space.
The semiclassical simulation reproduces the overall transmission but deviate from exact results and fail to capture short-time plateaus that arise when regions of Wigner-function negativity cross the barrier.
With the Kerr nonlinearity, reflections from nonlinear boundaries drive interference into classically forbidden regions, an effect that is inaccessible to semiclassical approaches.
We find that these interference effects do not alter the maximum transmission probability, which is bounded by the initial positive-energy fraction and therefore already encoded in the phase-space structure of the Fock states.
Because Fock states cannot be faithfully represented within classical phase space, the transmission through a barrier reveals fundamental limitations of semiclassical approaches.
\end{abstract}


\maketitle

\section{Introduction}

    One of the most intriguing phenomena in quantum mechanics is the transmission of particles through potential barriers higher than their kinetic energy, a process known as quantum tunneling~\cite{merzbacher2002early}. Since the early days of quantum theory, tunneling has played a central role in explaining molecular spectra \cite{hund1927deutung}, electron emission from metals \cite{nordheim1928theorie}, and $\alpha$-decay in nuclei \cite{gamow1928quantum}. Currently, tunneling phenomena can be simulated on different experimental platforms, such as NMR systems with a few spin nuclei \cite{feng2013experimental}, superconducting circuits \cite{devoret1985measurements,Fiske} and Bose-Einstein Condensates with ultracold atoms \cite{ramos2020measurement}. On the theoretical side, tunneling has been studied extensively, especially in double-well potentials \cite{zendra2024many,ferreira1997tunnelling,song2015localization,braun2008spontaneous,hasegawa2013gaussian}, but also in continuous-variable systems \cite{PradoKerr,marthaler2007quantum}, spin models \cite{Nader2021PRE,chen2006unconventional}, and Bose–Einstein condensates \cite{milburn1997quantum,zollner2008tunneling}.

    Many features of tunneling—such as wavepacket transmission and energy-level splitting—can be accurately captured using semiclassical approximations \cite{kluksdahl1988wigner,Kenneth,Jensen,tautermann2002accurate,rastelli2012semiclassical,aragon2020effective,LeDeunff}. However, semiclassical approaches cannot account for Wigner function negativity \cite{hillery1984distribution}, a genuine quantum feature with no classical analogue. From a practical perspective, Wigner negativity has been proposed as a necessary ingredient for outperforming classical devices \cite{lu2021propagating,rahimi2016sufficient,xiang2022distribution}. It is well established that the fine structure of the Wigner function—particularly
its negativity and interference fringes—is closely connected to the sensitivity of
quantum states to rotations and displacements~\cite{Toscano2006}. Despite growing interest, the precise role of Wigner negativity in tunneling and transmission phenomena, and its broader implications for nonclassical physics, remains under discussion~\cite{Lin2020,Chen2024,valtierra2024tunneling,ono2022phase}. In this context, marginal distributions are especially relevant for studying quantum barrier transmission, including contributions that may be associated with tunneling, since they correspond to directly measurable observables accessible through modern quantum state reconstruction techniques~\cite{Breitenbach,LEONHARDT199589,Lvovsky,Park2017}.

In this work, we examine to what extent the transmission effect can be captured
semiclassically using phase-space distributions. To this end,
we compare two approaches: the exact Wigner function and the Truncated Wigner Approximation
(TWA)~\cite{POLKOVNIKOV_TCA,TWA_2015,TWA,Corney2015}, a semiclassical method that, by
construction, neglects negative values. Because Fock states exhibit pronounced
Wigner-function negativity~\cite{kenfack2004negativity}, they provide natural test
cases—unlike Gaussian states~\cite{wang2007quantum} with strictly positive Wigner
functions—for examining the differences between classical and quantum phase-space
descriptions and their consequences for the transmission phenomenon. Our analysis starts with displaced Fock states evolving in an inverted oscillator, where the initial Wigner negativity remains constant due to the absence of nonlinear interactions. We then extend the study to the inverted oscillator with Kerr nonlinearity~\cite{Kerr},
a system of significant experimental relevance~\cite{marti2024kerr,he2023fast}, which
offers a viable route for testing the findings on currently available platforms.

section{Models and Hamiltonians}

In this section, we introduce two models for studying the transmission of displaced Fock states. The first is the inverted harmonic oscillator, which admits an analytical treatment, while the second extends it by including a small Kerr nonlinearity. The former serves as a fundamental model, whereas the latter is experimentally more accessible and incorporates nonlinear interactions.

\subsection{Inverted Harmonic Oscillator}

The inverted oscillator (IO) \cite{barton1986quantum} provides the simplest example of a potential barrier \cite{barton1986quantum,Yuce_2021,ullinger2022logarithmic}. Its Hamiltonian is given by
\begin{equation}
\label{Hio}
\hat{H} = \frac{1}{2}(\hat{p}^2 - \hat{q}^2) = -\frac{1}{2}(\hat{a}^{\dagger 2} + \hat{a}^2),
\end{equation}
where $\hat{q}$ and $\hat{p}$ are the canonical position and momentum operators, and $\hat{a}, \hat{a}^\dagger$ are the usual bosonic annihilation and creation operators.

 In classical mechanics, the inverted oscillator potential has a maximum at energy $E=0$. Particles with $E<0$ are reflected, while those with $E>0$ pass over the barrier \cite{ullinger2022logarithmic}. In quantum mechanics, the inverted oscillator generates squeezing \cite{albrecht1994inflation} and has been used to investigate exponential delocalization \cite{wang2023classicalquantum,romatschke2024out} as well as quantum–classical correspondence \cite{maamache2017quantum}. Analyzing the dynamics of quantum states in the inverted oscillator is particularly challenging because the associated phase space is unbounded \cite{maamache2017quantum}.

 In this work, the dynamics is initialized with displaced Fock states $\hat{D}(\alpha)|n\rangle$, where the displacement operator is defined as $\hat{D}(\alpha)={\rm exp}(\alpha\hat{a}^\dagger-\alpha^*\hat{a})$ and the displacement parameter as $\alpha=\frac{1}{\sqrt{2}}(\bar{q}+i\bar{p})$. A key property for our analysis is that the mean energy
\begin{eqnarray}
\label{meanenergyIO}
\bar{E}&=&\langle n| \hat{D}^\dagger(\alpha)\hat{H}\hat{D}(\alpha)|n\rangle\nonumber \\ &=&-(|{\rm Re}(\alpha)|^2-|{\rm Im}(\alpha)|^2),
\end{eqnarray}
is independent of the Fock index $n$ and only depends on the displacement parameter $\alpha$. This follows directly from the operator identities \cite{walls2012quantumoptics}  $\hat{D}^\dagger(\alpha)\hat{a}\hat{D}(\alpha)=\hat{a}+\alpha$ and $\hat{D}^\dagger(\alpha)\hat{a}\hat{D}(\alpha)=\hat{a}^\dagger+\alpha^*$. Consequently, the mean energy (\ref{meanenergyIO}) can be tuned below the barrier top ($E_c=0$) by choosing the displacement parameter such that  $|{\rm Im}(\alpha)|<|{\rm Re}(\alpha)|$.

\subsection{Inverted oscillator with a Kerr nonlinearity}

To address experimentally relevant scenarios, we extend our analysis to the inverted oscillator with a Kerr nonlinearity \cite{puri2017engineering,Kerr,chavez2023spectral}. The Hamiltonian is the following
\begin{equation}
\label{HKerr}
\hat{H} = -\epsilon_2(\hat{a}^{\dagger 2} + \hat{a}^2) + K \hat{a}^{\dagger 2} \hat{a}^2,
\end{equation}
where $\epsilon_2$ denotes a parametric driving strength and $K$ denotes the Kerr nonlinearity. This model is experimentally available with mechanical oscillators \cite{marti2024kerr}, superconducting
circuits\cite{he2023fast,cai2025quantum,venkatraman2024driven} and atoms in microcavities. \cite{rebic1999large}. On the theoretical side, it has been employed to study quantum tunneling \cite{PradoKerr,Kerr}, quantum phase transitions \cite{reynoso2025phase}, quantum-state engineering \cite{puri2017engineering,grimm2020stabilization}, and a broad range of other quantum phenomena \cite{chavez2023spectral,botin2017,iachello2023symmetries,holmes2023husimi,marthaler2006switching,Oliveira2006}.

In the semiclassical limit, we express the Hamiltonian in terms of canonical variables $(q, p)$
\begin{equation}
\label{Hsemiclassical}
H = \epsilon_2 (p^2 - q^2) + \frac{K}{4} (q^2 + p^2)^2.
\end{equation}
For $K=0$ and
$\epsilon_2=1/2$, the Hamiltonian (\ref{Hsemiclassical}) reduces to the inverted oscillator (\ref{Hio}). For $K>0$, the nonlinearity generates effective outer walls, and the resulting phase-space structure resembles that of a double-well potential with a separatrix at the critical energy $(E_c=0)$.

\begin{figure}[h]
\begin{center}
\begin{tabularx}{\textwidth}{cc}
  \parbox[t][2mm]{3mm}{\rotatebox[origin=c]{90}{\hspace{4cm}$p$}} &     \includegraphics[width=0.65\linewidth]{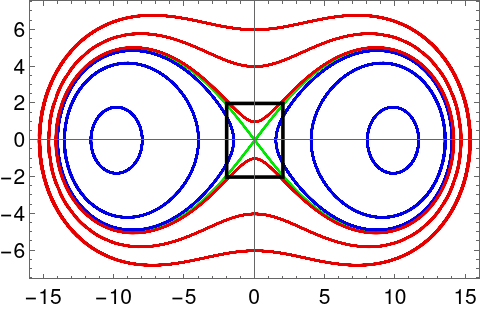}  \\
  \parbox[t][2mm]{3mm}{\rotatebox[origin=c]{90}{\hspace{4cm}$p$}} & \includegraphics[width=0.65\linewidth]{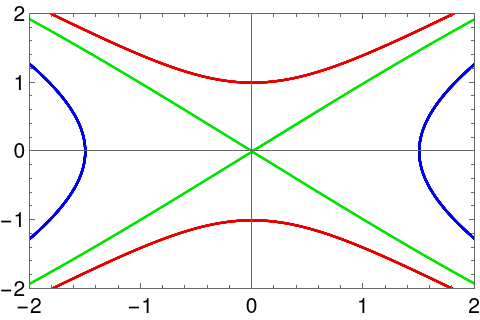}\\
  & \hspace{0.25cm}$q$
    \end{tabularx}
    \caption{\label{Trajectories} Classical trajectories of the  inverted oscillator with a small Kerr nonlinearity (\ref{Hsemiclassical}). The configuration of parameters is $\epsilon_2=1/2$ and $K=10^{-2}$. The green trajectory is the separatrix at the critical energy ($E_c=0$) corresponding to the top of the inverted oscillator.
    The blue/red trajectories correspond to energies smaller/larger than the top of the inverted oscillator ($E_c=0$).
    The bottom panel is a zoom into the center of the phase space in the top panel delimited by the square. At this small portion of the phase space the trajectories resemble those of the inverted oscillator\cite{ullinger2022logarithmic}.
    }
    \end{center}
\end{figure}

Figure~\ref{Trajectories} shows the classical trajectories in phase space, including the separatrix (green) at the critical energy
$E_c=0$ \cite{nader2025critical,NADER2023PLA}, sub-barrier trajectories (blue), and super-barrier trajectories (red). Near the origin, the trajectories closely resemble those of the inverted oscillator \cite{ullinger2022logarithmic}, providing a convenient baseline for comparison.

In the presence of the Kerr nonlinearity, the mean energy of displaced Fock states
\begin{equation}
\label{meanenergKerr}
\bar{E} = -(|{\rm Re}(\alpha)|^2-|{\rm Im}(\alpha)|^2) + K(|\alpha|^4 + n(n-1)  + 4n|\alpha|^2)\,,
\end{equation}
depends explicitly on the Fock index $n$. To maintain the initial states at the same energy below the threshold $E_c=0$, the displacement parameter $\alpha$ must therefore be adjusted as a function of $n$.

\section{Quantum Dynamics in the Phase Space}

We focus on the dynamics of displaced Fock states, $|\psi_0\rangle = D(\alpha)|n\rangle$ in the inverted oscillator with and without Kerr nonlinearity.
The preparation of such states is currently feasible with atoms in cavities \cite{Haroche2008} and in superconducting circuits \cite{Hofheinz}.
 For simplicity, we consider initial states localized on the left side of the inverted oscillator, with displacement parameter ${\rm Re}(\alpha)<0$ and mean energy smaller than the top of the inverted oscillator $\bar{E}<0$.

 To investigate the dynamics in phase space, we employ two complementary methods: A) the exact Wigner function and B) the Truncated Wigner Approximation (TWA). Both approaches are described in detail below.

\subsection{Quantum approach: Wigner function}

The Wigner function is a quasi-probability distribution that provides a phase-space representation of quantum states. It is defined as \cite{wignerfunctions,hillery1984distribution}
\begin{equation}
\label{Wigner}
W(q,p,t)=\frac{1}{\pi}\int \langle q+y |\psi(t)\rangle \langle \psi(t) | q-y \rangle  e^{2ipy}  d y \,,
\end{equation}
where $|\psi(t)\rangle$ is the wave function.  By integrating over momentum, the Wigner function yields the correct marginal probability distribution \cite{colomes2015comparing}:
\begin{equation}
\label{marginalP}
P(q,t) = \int W(q,p,t),dp = |\Psi(q,t)|^2 \,.
\end{equation}
 Since the Wigner function (\ref{Wigner}) can take negative values, it is not a true probability distribution in the classical sense  \cite{wignerfunctions}.  However, this negativity is not a flaw, but rather a fundamental feature that reflects quantum interference in phase space\cite{teske2018mean}. It is often interpreted as a signature of nonclassicality or quantum coherence \cite{kenfack2004negativity} and it is typically quantified as $\delta(t)=\int(|W(q,pt,t)|-W(q,p,t)) dxdp$.

For closed systems with initial pure states, the unitary evolution of the wave function is given by
$$\psi(t)=\hat{U}(t)|\psi_0\rangle\,,\,\,\,\hat{U}=\exp\left(-\frac{i}{\hbar}\hat{H}t\right).$$
 Some particular cases of interest are the class of coherent states $|\alpha\rangle= D(\alpha)|0\rangle$  whose Wigner functions are Gaussian\cite{adesso2014continuous} and everywhere positive, allowing them to be interpreted as classical-like probability distributions.
The transmission probability for a particle to be found in the  region $\Omega$ can be obtained through marginal probability (\ref{marginalP}) as
\begin{equation}
\label{probW}
P(\Omega,t)=\int_{-\infty}^{\infty}\int_{q_1}^{q_2} W(q,p,t) dq dp\,,
\end{equation}
where $q_1$ and $q_2$ are the boundaries of the region $\Omega$. For concreteness, we focus on states initially displaced to the left of the inverted oscillator and evaluate the probability of finding the particle on the right side, corresponding to  $q_1=0$ and $q_2=\infty$.

In the case of the pure inverted oscillator (\ref{Hio})—in the absence of nonlinear forces—the negativity of the initial state remains constant under time evolution \cite{gardiner2004quantum} $\delta(t)=\delta(t=0)$. This makes the inverted oscillator an ideal model to explore how the initial state's nonclassical features influence the transmission probabilities.  However, the quantum states in this system undergo exponential delocalization over time  \cite{wang2023classicalquantum}, in addition to the unbound phase space that prevents the numerical convergence of the integral (\ref{probW}) at large times. In contrast, the Kerr nonlinearity (\ref{HKerr}) represents a nonlinear force
that effectively bounds the phase space by creating effective outer walls. This confinement
enables stable numerical evaluation of the transmission probabilities (\ref{probW})
at longer times. Nevertheless, the nonlinear terms in the Kerr Hamiltonian (\ref{HKerr}) modify the negativity of the initial state $\delta(t)\neq \delta(t=0)$, leading to the emergence of interference patterns during the evolution, even for initially Gaussian states, whose Wigner functions are positive everywhere at
$t=0$.

 To quantify how much of each initial state lies above the top of the inverted harmonic potential,
we introduce a semiclassical measure of the population in the region of positive classical energy,
\begin{equation}
\label{probE}
P_0(E > 0) = \iint \Theta\!\big(H(q,p) - E_c\big)\, W(q,p,t=0)\, dq\,dp ,
\end{equation}
where \(E_c = 0\) and \(\Theta(x)\) is the Heaviside step function,
\begin{equation}
\Theta(x) =
\begin{cases}
0, & x < 0 ,\\[4pt]
1, & x \ge 0 .
\end{cases}
\end{equation}

The quantity \(P_0(E > 0)\) represents the positive-energy fraction of the initial Wigner distribution. However, (\ref{probE}) should not be interpreted as an exact quantum probability,
because the classical step function \(\Theta(H(q,p))\) does not coincide with the Weyl symbol of the corresponding quantum projector \(\Theta(\hat{H})\)
(see, e.g., \cite{hillery1984distribution,wignerfunctions}).
Nevertheless, it provides a convenient semiclassical reference for estimating the fraction of the initial Wigner distribution located in a different region of trajectories in the phase space (see Figure \ref{Trajectories}).

\subsection{Semiclassical approach: TWA}

Alternatively, the dynamics of quantum states can be simulated using the truncated Wigner approximation (TWA) \cite{TWA_2015,POLKOVNIKOV_TCA,Corney2015}. This semiclassical method leverages the stochastic nature of quantum mechanics by generating an ensemble of $N$ points in classical phase space that collectively approximate the Wigner function of the initial quantum state.  Each point in this ensemble evolves independently according to Hamilton’s equations of motion
\begin{equation}
\label{canonical}
\dot{q}=\frac{\partial H}{\partial p}\,,\,\,\,\,\dot{p}=-\frac{\partial H}{\partial q}\,,
\end{equation}
where $H$ denotes the classical Hamiltonian. For initial coherent states $|\psi_0\rangle=\hat{D}(\alpha)|0\rangle$, the Wigner function is strictly positive and can be directly interpreted as a classical probability distribution. In contrast, for displaced Fock states $D(\alpha)|n\rangle$,  which exhibit Wigner negativity, an additional approximation is required to generate a positive-definite sampling distribution. To address this, we employ a Gaussian approximation \cite{PositiveP} for Fock states, defined as
\begin{equation}
\label{Ppositive}
P_n(q,p)=\sqrt{\frac{2}{\pi}}{\rm exp}\left( \frac{(q^2/2+p^2/2-n-1/2)^2}{1/2}\right)\,,
\end{equation}
where $n$ is the Fock state index. This distribution approximates the Wigner function of the initial displaced Fock state and is used to generate the initial ensemble for TWA simulations, as illustrated in Figure \ref{Fockshots}.

\begin{figure}[h]
\begin{center}
\begin{tabularx}{\textwidth}{ccc}
& \hspace{-0.5cm}Wigner & \hspace{-0.5cm} Gaussian approx \\
\multicolumn{3}{c}{$t=0$}\\
  \parbox[t][2mm]{3mm}{\rotatebox[origin=c]{90}{\hspace{3.5cm}$p$}} &     \includegraphics[width=0.4\linewidth]{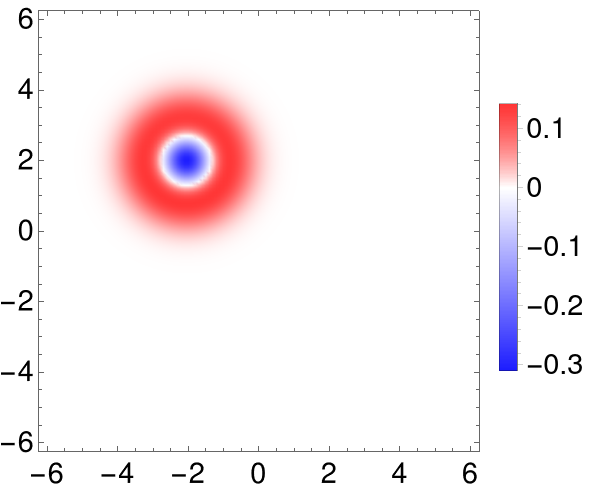} & \includegraphics[width=0.4\linewidth]{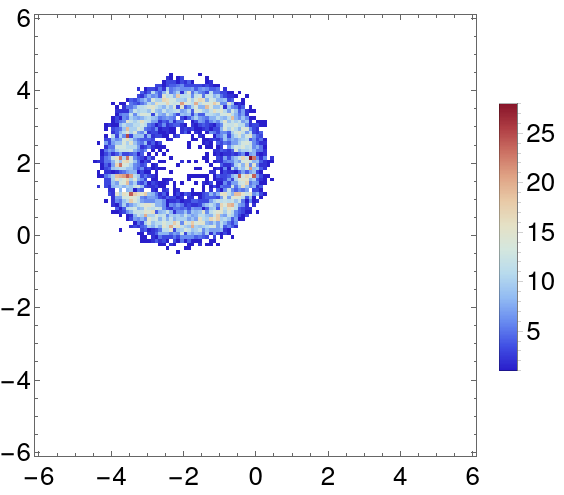}  \\
  & \hspace{-0.5cm}Wigner & \hspace{-0.5cm} Gaussian approx + TWA \\
  \multicolumn{3}{c}{$t=1$}\\
  \parbox[t][2mm]{3mm}{\rotatebox[origin=c]{90}{\hspace{3.5cm}$p$}} & \includegraphics[width=0.4\linewidth]{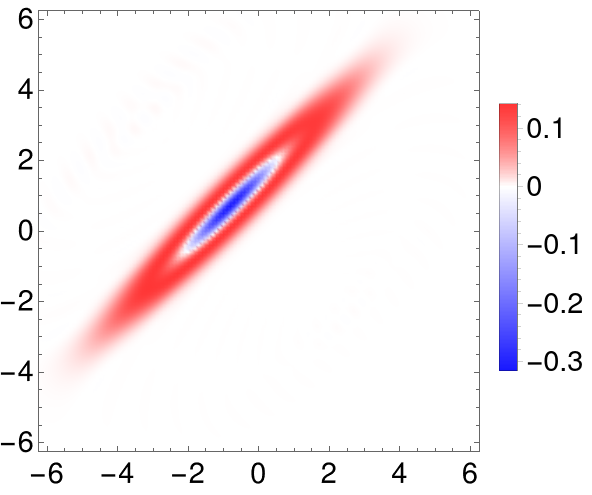} & \includegraphics[width=0.4\linewidth]{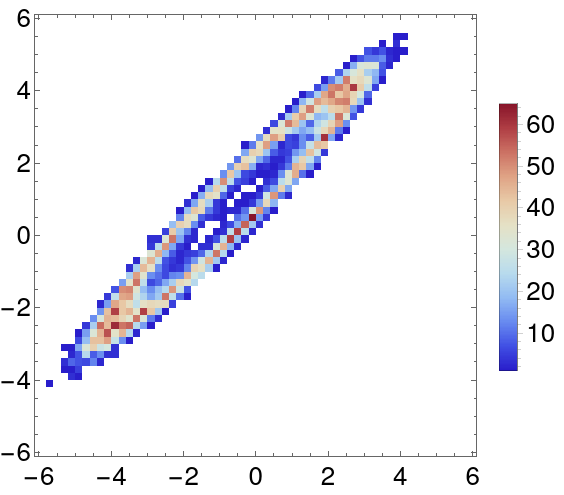} \\
  & \hspace{-0.5cm}$q$ & \hspace{-0.5cm}$q$
    \end{tabularx}
    \caption{\label{Fockshots}  (Left column) Displaced Fock state $D(\alpha)|1\rangle$ with displacement parameter $\alpha=\frac{1}{2}(-2+2i)$ at different times evolving in the inverted oscillator (\ref{Hio}). (Right column) Histogram of the set of $N=10^5$ points that replicate the Fock state according to the Gaussian approximation (\ref{Ppositive}).
    }
    \end{center}
\end{figure}

Within the TWA framework, only those phase-space points ($q_i,p_i$), with positive energy $E(q_i,p_i)>0$ exceeding the top of the inverted oscillator, can contribute to the transmission probability into the region $\Omega$ delimited by the turning points $q_{1,2}$. The transmission probability is then estimated using the Heaviside step function
\begin{equation}
\label{probTWA}
P(\Omega,t)=\frac{1}{N}\sum_{i=1}^N\Theta(q_i(t)-q_1)-\frac{1}{N}\sum_{i=1}^N\Theta(q_i(t)-q_2)\,.
\end{equation}

This approach is particularly advantageous for avoiding the direct evaluation of the Wigner function integral in  (\ref{probW}), especially in regimes where numerical convergence is challenging. However, by construction, the TWA does not account for negative regions of the Wigner function and therefore cannot capture quantum interference effects associated with Wigner negativity.

\section{Results}

\subsection{Coherent states in the inverted oscillator}

For the specific case of a coherent state evolving under the inverted harmonic oscillator, the  transmission probability given in (\ref{probW}) can be evaluated analytically. In the absence of nonlinear forces, the dynamics of the Wigner function is entirely classical, in the sense that it obeys the Liouville equation \cite{wignerfunctions}
\begin{equation}
\label{Moyal}
\frac{\partial W}{\partial t}=-p\frac{\partial W}{\partial q}+\frac{\partial U}{\partial q}\frac{\partial W}{\partial p}\,.
\end{equation}
The solution at a later time $t$ is obtained by following the classical trajectories backward in time
$W(q,p,t)=W(q_{-t}(q,p),p_{-t}(q,p),0)$
where $x_t$ and $p_t$ are the classical trajectories governed by Hamilton’s equations of motion (\ref{canonical}). For the inverted oscillator, these trajectories take the form \cite{Koh:2023qjq,bagrov2013coherent}
\begin{eqnarray}
\label{eqsmovIHO}
q_t&=&q\cosh(t) + p \sinh( t) \nonumber \\
p_t&=&p\cosh(t) + q \sinh( t)\,.
\end{eqnarray}

Using these expressions, the time-evolved Wigner function of an initial coherent state centered at $(q_0,p_0)$ becomes
\begin{equation}
\label{Wcoherentt}
W(q,p,t)=\frac{1}{\pi}e^{-(q\cosh(-t)+p\sinh(-t)-q_0)^2}e^{-(p\cosh(-t)+q\sinh(-t)-p_0)^2}\,.
\end{equation}
 Substituting  (\ref{Wcoherentt}) into (\ref{probW})  yields an integral that can be solved analytically using the error function
\cite{error}. The resulting expression for the transmission probability is
\begin{eqnarray}
\label{Panalytical}
P(q>0,t) &=&\frac{1}{2} \left( 1 - {\rm erf}(-c(t) \right)\,,
\end{eqnarray}
where $c(t)=\frac{q_0\cosh(t)+p_0\sinh(t)}{\sqrt{\cosh^2(t)+\sinh^2(t)}}$.

\begin{figure}[h]
\begin{center}
\begin{tabularx}{\textwidth}{cc}
& $|\psi_0\rangle=\hat{D}(\alpha)|0\rangle$ \\
  \parbox[t][2mm]{3mm}{\rotatebox[origin=c]{90}{\hspace{2.5cm}$P(q>0,t)$}} & \includegraphics[width=0.65\linewidth]{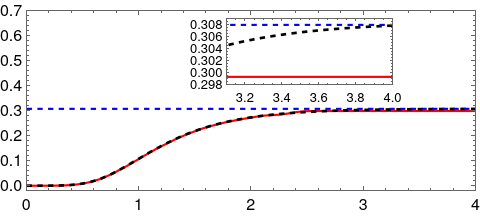}\\
  & $t$
    \end{tabularx}
    \caption{\label{Pcoherent} Probability $P(q>0,t)$ of finding the particle in the right side of the inverted oscillator as a function of the time. The solid red line corresponds to the TWA and the black dashed line to the exact analytical formula (\ref{Panalytical}). The blue dashed line represents the probability  $P_0(E>0)=0.308$. The initial state is a coherent state $|\psi_0\rangle=\hat{D}|0\rangle$ with displacement parameter $\alpha=\frac{1}{\sqrt{2}}(-3+2.5i)$.
    }
    \end{center}
\end{figure}

Figure~\ref{Pcoherent} shows the time evolution of the transmission probability $P(q>0,t)$ for an initial coherent state with displacement parameter $\alpha=\frac{1}{\sqrt{2}}(-3+2.5i)$. The probability increases smoothly over time and asymptotically approaches roughly $P_0(E>0)=0.308$. The transmission probability from the TWA (\ref{probTWA}) is found to be in good agreement with the exact analytical result (\ref{Panalytical}). At $t=4$, the exact transmission probability $P_{ex}(x>0,t=4)=0.308$ while the TWA gives a slightly lower value $P_{TWA}(x>0,t=4)=0.299$.
This correspondence is expected, since in the absence of nonlinear forces the state remains Gaussian throughout its evolution \cite{wignerfunctions}, and its dynamics can be accurately captured by classical phase space trajectories.

\subsection{Displaced Fock states in the inverted oscillator}

More interesting are the initial displaced Fock states $|\psi_0\rangle=|\hat{D}(\alpha)|n\rangle$, with Fock index $n\neq0$, which exhibit significant Wigner function negativity.  For these non-Gaussian states, the transmission probability (\ref{probW}) must be evaluated numerically, and can be computed
with the target accuracy of $10^{-6}$ only for short times $t < 1.6$, due to the increasing delocalization of the Wigner function.

\begin{figure}[h]
\begin{center}
\begin{tabularx}{\textwidth}{cc}
& $|\psi_0\rangle=\hat{D}(\alpha)|1\rangle$ \\
  \parbox[t][2mm]{3mm}{\rotatebox[origin=c]{90}{\hspace{4cm}$P(q>0,t)$}} & \includegraphics[width=0.65\linewidth]{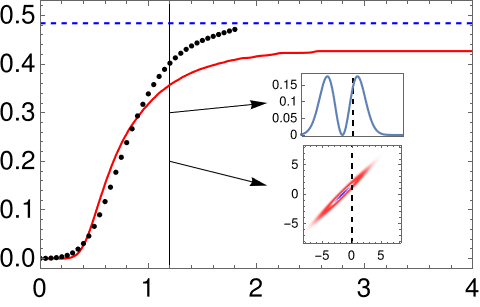}\\
  & $|\psi_0\rangle=\hat{D}(\alpha)|2\rangle$ \\
  \parbox[t][2mm]{3mm}{\rotatebox[origin=c]{90}{\hspace{4cm}$P(q>0,t)$}} & \includegraphics[width=0.65\linewidth]{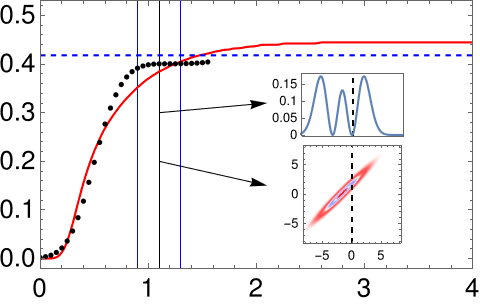} \\
   & $|\psi_0\rangle=\hat{D}(\alpha)|3\rangle$ \\
  \parbox[t][2mm]{3mm}{\rotatebox[origin=c]{90}{\hspace{4cm}$P(q>0,t)$}} & \includegraphics[width=0.65\linewidth]{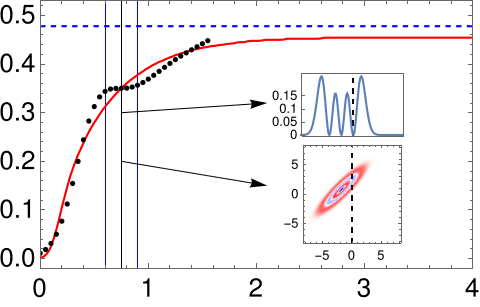} \\
  & $t$
    \end{tabularx}
    \caption{\label{Pfock}  Probability $P(q>0,t)$ of finding the particle at the right side of the inverted oscillator. The initial states are the displaced Fock state  $D(\alpha)|n\rangle$ with displacement parameter $\alpha=\frac{1}{\sqrt{2}}(-3+2.5i)$. In the inverted oscillator (\ref{Hio}) all states have the same mean energy $\bar{E}=-1.375$. The solid red curve shows the transmission probability from the Gaussian approximation + TWA method  (\ref{probTWA}), while the black points show the exact marginal probability (\ref{probW}). The blue dashed line represents the probability $P_0(E>0)$. The blue vertical lines indicate the start and end of the short-time plateaus. The insets show the Wigner function (\ref{Wigner}) and the marginal distribution (\ref{marginalP}) at the time indicated by  the vertical black line.
    }
    \end{center}
\end{figure}

Figure~\ref{Pfock} shows the time evolution of the transmission probability
$P(q>0,t)$ for different initial displaced Fock states. According to
(\ref{meanenergyIO}), displaced Fock states with the same displacement
parameter share the same mean energy. For the analysis we chose
$\alpha = \tfrac{1}{\sqrt{2}}(-3+2.5i)$. The agreement between the exact
numerical results and the TWA method is good at short times. However, the exact
dynamics display short-time plateaus in the tunneling probability that are absent
in the TWA. The instantaneous Wigner functions shown in
the insets of Figure \ref{Pfock} reveal that these plateaus arise when the negative
regions of the Wigner function cross into the right side of the inverted oscillator.
The corresponding marginal distributions indicate that these features are associated  with zeros of the probability distribution crossing the inverted oscillator. This effect cannot be captured within a classical phase-space description.

Although there are indications that the transmission probabilities obtained from the exact Wigner
dynamics and from the TWA may converge to different asymptotic values, this cannot be confirmed within the present
model because the rapid delocalization of the states prevents the convergence of the transmission
integral (\ref{probW}). In all cases, however, the exact transmission probability
\(P(x>0,t)\) does not exceed the initial positive-energy fraction \(P_0(E>0)\), contrary to the results from the TWA. When the Kerr
nonlinearity is included, the phase space becomes effectively bounded, and the integrals defining
the asymptotic probabilities are expected to converge, allowing for more reliable quantitative
comparisons.

\subsection{Displaced Fock states in the inverted oscillator with a small Kerr nonlinearity}

The analysis of the time evolution of displaced states can be extended to the inverted oscillator with a small Kerr nonlinearity (\ref{HKerr}). In this case, the Kerr nonlinearity
$K$ introduces effective outer boundaries in phase space, together with nonlinear forces that can distort the negative regions of the Wigner function. For small values of
$K$, the dynamics remain close to those of the inverted harmonic oscillator over short time scales. At longer times, however, the effective boundaries drive the state along the separatrix and eventually produce a bouncing effect.

As an example, Figure \ref{Fockshots} shows the exact Wigner function and the corresponding Gaussian approximation + TWA simulation at selected times for the initial displaced Fock state
$\hat{D}(\alpha)\lvert 1\rangle$. For short times ($t \lesssim 1$), the state is
squeezed, just as it is in the pure inverted oscillator (see Figure \ref{Fockshots}). At longer times,
however, the state follows the curvature of the separatrix of the inverted Kerr oscillator,
reflecting the presence of effective boundaries imposed by the Kerr nonlinearity.
The exact Wigner function penetrates into the right-hand side of the separatrix,
$\Omega_r$, giving rise to interference patterns. The interference in the right side of the separatrix
corresponds to transmission through classically forbidden regions, while those in the left side may be associated with partial reflection of above-barrier components.
In contrast, the TWA simulations cannot access the classically forbidden region
$\Omega_r$, since the semiclassical trajectories remain disconnected (see Figure \ref{Trajectories}).
This penetration into the classically forbidden region is therefore a genuinely quantum
phenomenon that cannot be reproduced within any classical phase-space description.

\begin{figure}[h]
\begin{center}
\begin{tabularx}{\textwidth}{ccc}
& \hspace{-0.5cm}Wigner & \hspace{-0.5cm} TWA
 \\
\multicolumn{3}{c}{$t=0$}\\
  \parbox[t][2mm]{3mm}{\rotatebox[origin=c]{90}{\hspace{3.5cm}$p$}} &     \includegraphics[width=0.4\linewidth]{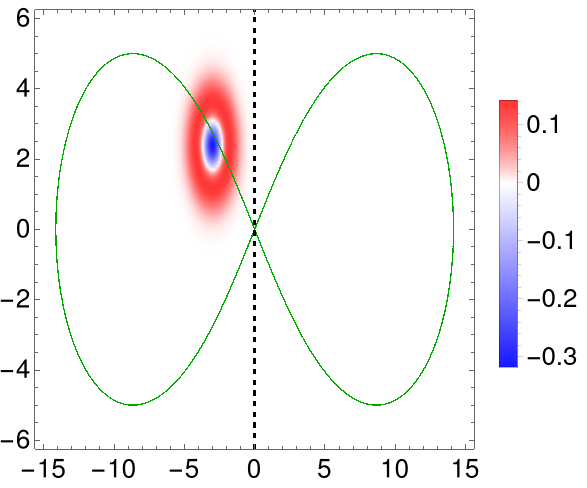} & \includegraphics[width=0.4\linewidth]{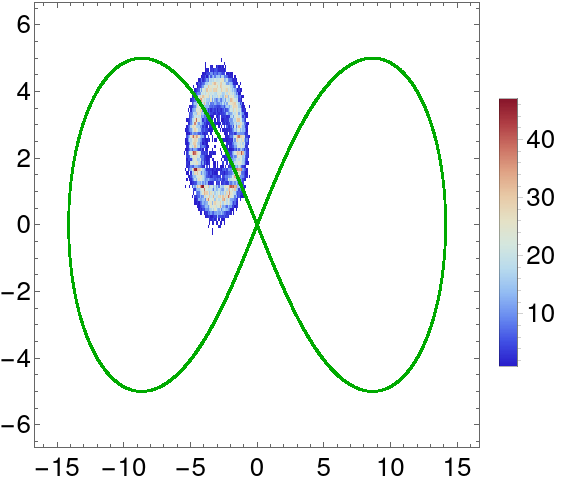}  \\
\multicolumn{3}{c}{$t=1$}\\
  \parbox[t][2mm]{3mm}{\rotatebox[origin=c]{90}{\hspace{3.5cm}$p$}} &     \includegraphics[width=0.4\linewidth]{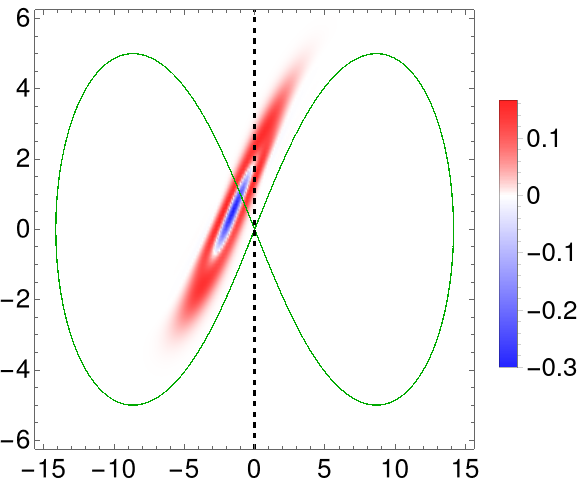} & \includegraphics[width=0.4\linewidth]{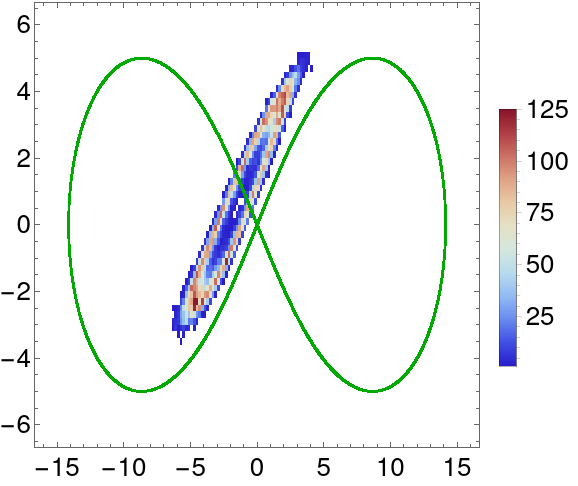}  \\
\multicolumn{3}{c}{$t=2$}\\
  \parbox[t][2mm]{3mm}{\rotatebox[origin=c]{90}{\hspace{3.5cm}$p$}} &     \includegraphics[width=0.4\linewidth]{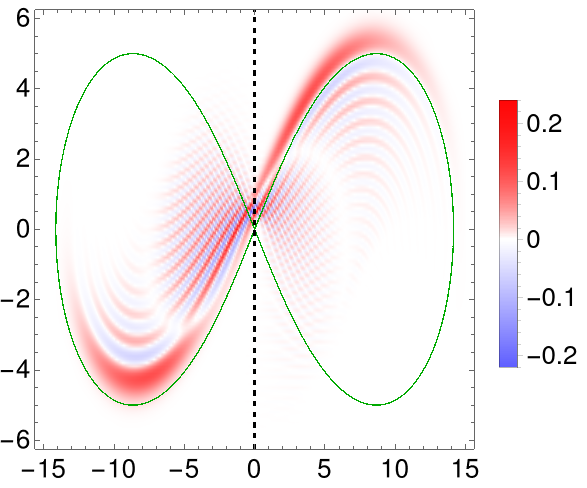} & \includegraphics[width=0.4\linewidth]{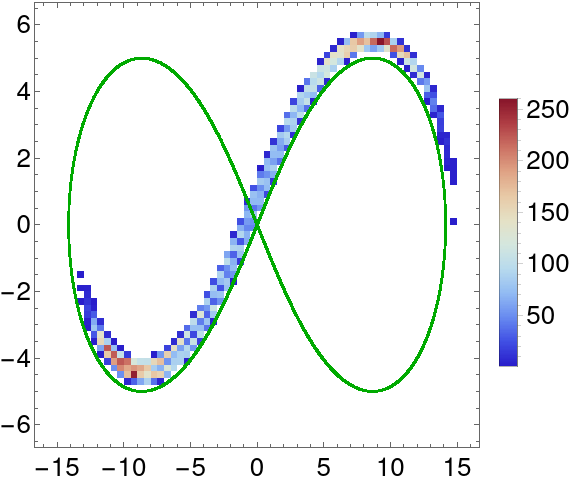}  \\
  & \hspace{-0.5cm}$q$ & \hspace{-0.5cm}$q$
    \end{tabularx}
    \caption{\label{Fockshots}  (Left column) Dynamics of the Wigner function of the initial displaced state $D(\alpha)|1\rangle$ with displacement parameter $\alpha=\frac{1}{\sqrt{2}}(-3+2.395i)$ in the Kerr inverted oscillator. On the left column the exact Wigner function and on the right column the TWA simulations with $N=10^4$ points. The green line is the separatrix. For all figures the driving strength $\epsilon_2=1/2$ and the Kerr non-linearity $K=10^{-2}$ are fixed. The right side of the separatrix, $\Omega_r$, is a classically forbidden region that is inaccessible to the TWA method.
    }
    \end{center}
\end{figure}

\begin{figure}[h]
\begin{center}
\begin{tabularx}{\textwidth}{cc}
& $|\psi_0\rangle=\hat{D}(\alpha)|0\rangle$ with $\alpha=\frac{1}{\sqrt{2}}(-3+2.5i)$ \\
  \parbox[t][2mm]{3mm}{\rotatebox[origin=c]{90}{\hspace{4cm}$P(q>0,t)$}} & \includegraphics[width=0.58\linewidth]{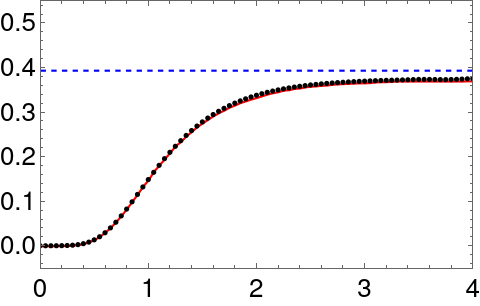}\\
& $|\psi_0\rangle=\hat{D}(\alpha)|1\rangle$ with $\alpha=\frac{1}{\sqrt{2}}(-3+2.395i)$ \\
  \parbox[t][2mm]{3mm}{\rotatebox[origin=c]{90}{\hspace{4cm}$P(q>0,t)$}} & \includegraphics[width=0.58\linewidth]{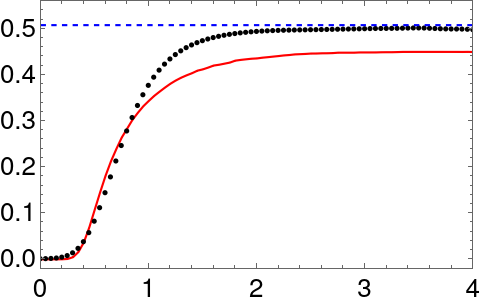}\\
  & $|\psi_0\rangle=\hat{D}(\alpha)|2\rangle$ with $\alpha=\frac{1}{\sqrt{2}}(-3+2.285i)$ \\
  \parbox[t][2mm]{3mm}{\rotatebox[origin=c]{90}{\hspace{4cm}$P(q>0,t)$}} & \includegraphics[width=0.58\linewidth]{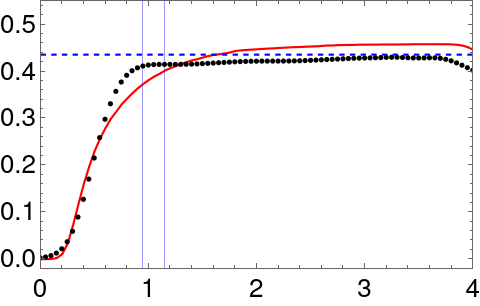}\\
    & $|\psi_0\rangle=\hat{D}(\alpha)|3\rangle$ with $\alpha=\frac{1}{\sqrt{2}}(-3+2.17i)$ \\
  \parbox[t][2mm]{3mm}{\rotatebox[origin=c]{90}{\hspace{4cm}$P(q>0,t)$}} & \includegraphics[width=0.58\linewidth]{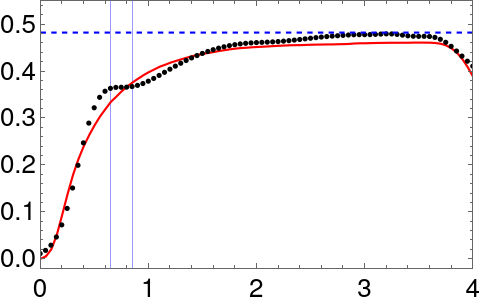}\\
  & $t$
    \end{tabularx}
\caption{\label{PKerr} Probability $P(q>0,t)$ of finding the particle on the right side of the Kerr
inverted oscillator, starting from displaced Fock states $\hat{D}(\alpha)|n\rangle$.
 All initial states have $\bar{q}=-3$ and mean energy $\bar{E}\approx -0.79$, below the
barrier threshold. The solid red curve shows the transmission probability from the TWA method (\ref{probTWA}), black points denote the exact marginal
probability (\ref{probW}), and the blue dashed line the probability $P_0(E>0)$. The vertical blue line indicates the onset and end of
the short-time plateaus. Parameters: $\epsilon=1/2$, $K=10^{-2}$. The analysis is
terminated at $t=3$, before the states return to the left side of the oscillator.     }
    \end{center}
\end{figure}

A natural question arising from the dynamics of the displaced Fock states shown in
Figure \ref{Fockshots} is whether the emergence of quantum interference modifies the behavior of the transmission probability over time.
Figure \ref{PKerr} displays the time evolution of the transmission probability \(P(q>0,t)\) for
several initial displaced Fock states \(D(\alpha)\lvert n\rangle\).
In the presence of the Kerr nonlinearity, the displacement parameter is adjusted so that all
states share the same mean energy, \(E \approx -0.79\).
Because the Kerr nonlinearity is weak, the dynamics closely resemble those of the inverted
oscillator, including the short-time plateaus that appear when regions of the Wigner function
negativity cross the barrier.
Owing to the bounded phase space, the integral defining the transmission probability
(\ref{probW}) converges at longer times, allowing for meaningful asymptotic comparisons.

For coherent states (\(n=0\)), the agreement between the exact and TWA transmission probabilities
remains remarkably good.
For displaced Fock states with \(n \neq 0\), the asymptotic transmission probabilities show
systematic but generally small deviations from the exact quantum results.
In most cases, the TWA underestimates the transmission probability,
although in certain instances it slightly overestimates it.

More importantly, it is confirmed that the exact transmission probability \(P(x>0,t)\) never exceeds the initial
positive-energy fraction \(P_0(E>0)\).
This result demonstrates that, although quantum interference develops in the classically forbidden
regions, the combined contribution of positive and negative interference patterns cancels out and the total transmission remains bounded by \(P_0(E>0)\). According to this result, the maximum possible value of the transmission probability is already encoded in the initial state.

With regards to the TWA simulations, it is the inability of the Gaussian approximation (\ref{Panalytical}) to accurately replicate the nonclassical initial state that ultimately leads to discrepancies in the predicted transmission probability.


\begin{figure}[h]
\begin{center}
\begin{tabularx}{\textwidth}{cc}
& (a) Transmission probability vs mean energy \\
  \parbox[t][2mm]{3mm}{\rotatebox[origin=c]{90}{\hspace{3cm}$P(q>0,t=3)$}} & \includegraphics[width=0.6\linewidth]{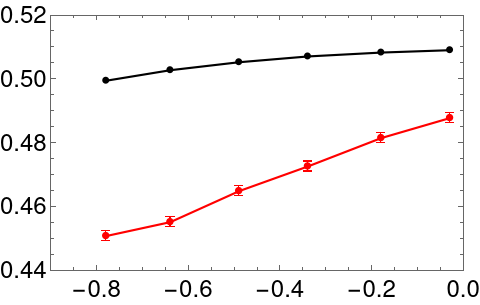}\\
 & $\bar{E}$ \\
  & (b) Transmission probability vs Fock index \\
  \parbox[t][2mm]{3mm}{\rotatebox[origin=c]{90}{\hspace{3cm}$P(q>0,t=3)$}} & \includegraphics[width=0.6\linewidth]{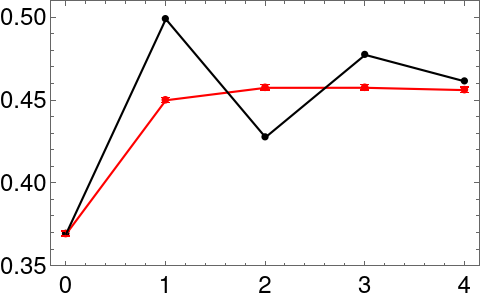}\\
  & $n$
    \end{tabularx}
\caption{\label{discrepancies} (a) Transmission probability $P(q>0,t=3)$ for the displaced Fock state
$\hat{D}(\alpha)\lvert 1\rangle$ in the inverted oscillator with $\epsilon_2 = 1/2$
and Kerr nonlinearity $K = 10^{-2}$, shown as a function of the mean energy.
The displacement parameter is
$\alpha = \tfrac{1}{\sqrt{2}}(-3+ip)$, with the imaginary part varied so that the
mean energy approaches the barrier top $E_c = 0$. The probability is evaluated at $t=3$, before the state reflects back to the left
side of the inverted oscillator.
(b) Transmission probability $P(q>0,t=3)$ for displaced Fock states
$\hat{D}(\alpha)\lvert n\rangle$ in the same system, shown as a function of the
Fock index $n$. All initial states are prepared with mean energy $\bar{E} = -0.79$
and mean position $\bar{q} = -3$. In both panels, The red points represents the transmission probability from the TWA method  (\ref{probTWA}), while the black points show the exact marginal probability (\ref{probW}). }
    \end{center}
\end{figure}

 Figure \ref{discrepancies} panel (a) shows the energy variance
$\sigma_H=\sqrt{\langle\hat{H}^2\rangle - \langle \hat{H}\rangle^2}$
of the initial displaced Fock states presented in Figure \ref{PKerr}.
The variance increases monotonically with the Fock index $n$, indicating that higher-order Fock states possess broader energy distributions.
In contrast, the corresponding positive-energy fraction $P_0(E>0)$,
shown in Figure \ref{discrepancies} panel (b), exhibits a nonmonotonic dependence on $n$.
This behavior arises because $P_0(E>0)$ is governed by the relative position
of the positive and negative lobes of the Wigner function with respect to the separatrix.
Consequently, while a larger energy variance generally favors transmission, the detailed interference pattern of each displaced Fock state also plays a crucial role.  The interplay between the energy spread and the fine structure of the Wigner function explains why the transmission probability does not simply scale with $n$ but instead exhibits an oscillatory dependence,
in which quantum interference fringes alternately enhance or degrade the transmission.
Similar behavior — namely, the modulation of transmission by quantum interference structures — has been observed in single-molecule transistors, where destructive interference between two conduction channels suppresses leakage and thereby enhances device performance \cite{chen2024quantum}.

\begin{figure}[h]
\begin{center}
\begin{tabularx}{\textwidth}{cccc}
& {\small (a) Energy variance} & & {\small (b) Positive-energy fraction} \\
  \parbox[t][2mm]{3mm}{\rotatebox[origin=c]{90}{\hspace{2.5cm}$\sigma_{H}$}} & \includegraphics[width=0.45\linewidth]{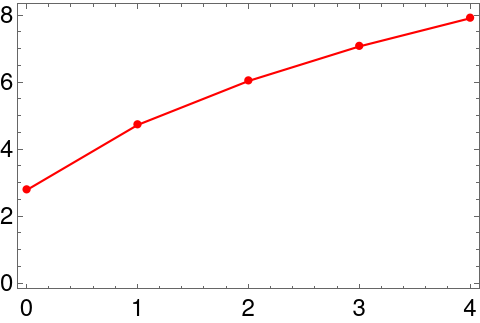} &
  \parbox[t][2mm]{3mm}{\rotatebox[origin=c]{90}{\hspace{2.5cm}$P_0(E>0)$}} & \includegraphics[width=0.45\linewidth]{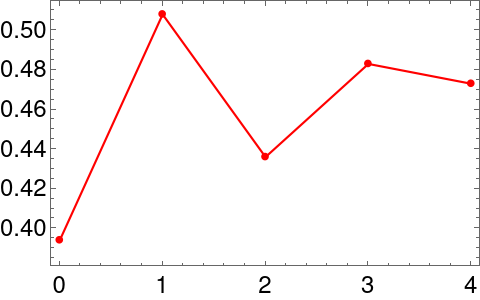}\\
  & $n$ & & $n$
    \end{tabularx}
\caption{\label{Var} (a) Variance $\sigma_H$ of the Hamiltonian (\ref{HKerr})  and (b) Positive-energy fraction $P_0(E>0)$ for the initial states in Figure \ref{PKerr} as a function of the Fock index. }
    \end{center}
\end{figure}

In Figure \ref{discrepancies} panel (a) we show that the discrepancy between the TWA and the exact quantum results decreases and eventually vanishes as the mean energy approaches the barrier top. Moreover, Figure  \ref{discrepancies} panel (b) illustrates that the agreement between the TWA and the exact results systematically improves with increasing Fock index
$n$ of the initial state. This trend is consistent with the expected validity of the Gaussian approximation \cite{PositiveP}. As the Fock index  increases, the Wigner function approaches a narrow, nearly Gaussian ring in phase space, thereby reducing the influence of interference fringes and enhancing the accuracy of the semiclassical description.

\begin{figure}[h]
\begin{center}
\begin{tabularx}{\textwidth}{cc}
& (a) Wigner volume \\
  \parbox[t][2mm]{3mm}{\rotatebox[origin=c]{90}{\hspace{3cm}$\int_{\Omega_r}|W(x,p,t)|dxdp$}} & \includegraphics[width=0.6\linewidth]{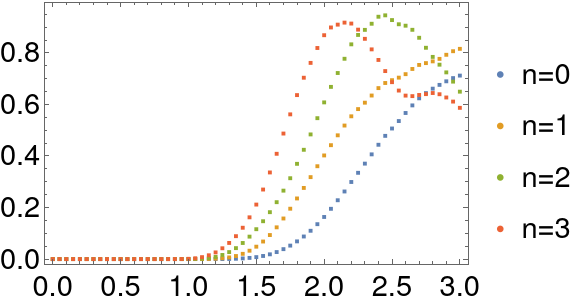}\\
  & (b) Fringes \\
  \parbox[t][2mm]{3mm}{\rotatebox[origin=c]{90}{\hspace{3cm}$N_s$}} & \includegraphics[width=0.6\linewidth]{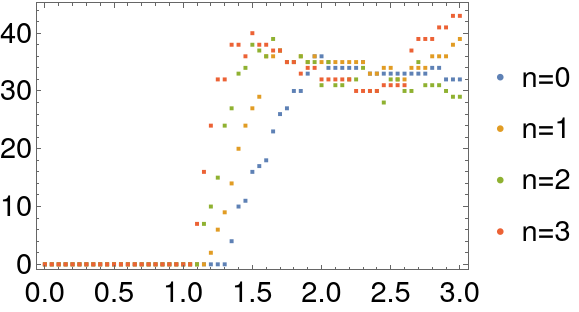}\\
  & $t$
    \end{tabularx}
\caption{\label{Wvolume} (a) Volume of the absolute Wigner function in the classically forbidden region $\Omega_r$ as a function of time.
(b) Number $N_s$ of sign changes in the slice of the Wigner function $W(x,0,t)$ along momentum and within the classically forbidden region $\Omega_r$. The number of sign changes quantify the fringe structure. The color of the points indicate the Fock index of the initial displaced Fock $|\psi_0\rangle = \hat{D}(\alpha)|n\rangle$ states. The displacement parameters are chosen to keep the mean energy fixed at $\bar{E}\approx-0.79$ (see Figure \ref{PKerr}). For all figures the driving strength $\epsilon_2=1/2$ and the Kerr non-linearity $K=10^{-2}$ are fixed. }
    \end{center}
\end{figure}

The nonzero volume of the Wigner function in the classically forbidden region
$\Omega_r$, highlights the
limitations of the classical simulation. Figure~\ref{Wvolume} shows the absolute
volume of the Wigner function in $\Omega_r$ as a function of time for different
initial Fock states. These results confirm that the ability of the states to
penetrate into the classically forbidden region increases with the initial Fock
index $n$. Furthermore, the number of sign changes observed in the slice of the Wigner function along $p=0$ quantify the interference pattern within the classically forbidden
region.

\section{Conclusions}

  We have investigated the transmission of Fock states through an inverted oscillator, both with and without Kerr nonlinearity. For the analysis we concentrate on states with average energies below the top of the inverted oscillator. It was used a Gaussian approximation to replicate the Fock states in the classical phase space and the TWA to track the dynamics. The semiclassical results were compared against those from the Wigner function and the marginals probabilities. For coherent states, which have strictly positive Wigner functions, the transmission probability shows good agreement between the TWA and the exact marginal probabilities. In contrast, for displaced Fock states exhibiting substantial Wigner negativity, clear discrepancies arise even at short times. In particular, short-time plateaus emerge when negative regions of the Wigner function (or, equivalently, zeros of the probability distribution) cross the inverted oscillator barrier. These plateaus cannot be reproduced semiclassically and thus constitute a genuine quantum effect. At longer times in the inverted oscillator, the rapid delocalization of the states prevents convergence of the transmission probability, leaving the asymptotic comparisons inconclusive.

Introducing a Kerr nonlinearity effectively confines the phase space and enables the analysis to be extended to longer times. However, the Kerr nonlinearities also introduce forces that modify the negativity of the states. At short times the states are squeezed just as they are in the pure inverted oscillator. At longer times, the dynamics follow the curvature of the separatrix while interference fringes develop within the classically forbidden region. This is the second phenomenon impossible to simulate in the classical phase space by the TWA approximation since the trajectories are disconnected. The transmission probabilities from the TWA and the exact marginals exhibit systematic discrepancies that decrease as the index of the Fock state increase or as the average energy approaches the critical energy (the barrier top).

Finally, we find that the exact transmission probability $P(x>0,t)$ never exceeds
the initial positive-energy fraction $P_0(E>0)$.
This demonstrates that the maximum attainable transmission is already encoded
in the energetic structure of the initial Fock state.
Because Fock states cannot be faithfully replicated within a classical phase-space framework,
the semiclassical approximations remain effectively unaware of this bound.

\section{ Numerical detail and Data Availability}

All codes, scripts and data needed to reproduce the results in this manuscript are available online \cite{NaderIO}

\url{https://github.com/djuliannader/InvertedOscillator}

The exact Wigner functions (\ref{Wigner}) were computed using the
\textit{QuantumOptics.jl} framework~\cite{QuantumOptics}, and the integrals for
the transmission probabilities (\ref{probW}) were evaluated with the Julia
package \textit{HCubature}~\cite{HCubature}. The Fock basis was truncated at
$n_{\mathrm{max}} = 100$, and the relative error tolerance for the integrals was
set to $10^{-6}$. The integration domain was chosen as
$q \in [-15,15]$ and $p \in [-15,15]$.  As for the Gaussian approximation + TWA we used $N=10^5$ sampling points to replicate the initial states. The canonical equations (\ref{canonical}) were solved numerically using NDSolve from Mathematica \cite{NDSolve}.
The statistical error for the transmission probability (\ref{probTWA}) is estimated as usual for binomial distributions \cite{casella2024statistical}
\begin{equation}
\label{error}
\sigma=\sqrt{\frac{P(q>0,t)(1-P(q>0,t))}{N}}\,.
\end{equation}
Error bars are only visible at the scale of Figure \ref{discrepancies}. When a Kerr nonlinearity is introduced $K>0$, the Gaussian approximation systematically overestimates the mean energy. To compensate for this shift, we manually adjust the initial average momentum so that the states in the Gaussian approximation have the same mean energy as in the exact solution.

\section{Acknowledgments}
D.J.N. acknowledges financial support from the Programme Johannes Amos
Comenius under the Ministry of Education, Youth and Sports of the Czech
Republic reg. no. CZ.02.01.01/00/22\_008/0004649 D.J.N. thanks D. W. Moore and S. Lerma-Hern\'andez for useful discussions.

\bibliography{Bibliography.bib}

\end{document}